\documentclass[aps,prl,twocolumn,showpacs]{revtex4}
\usepackage{graphicx}
\usepackage{dcolumn}
\usepackage{bm}
\begin{document}
\title{Disordered Kondo Nanoclusters: Effect of Energy Spacing}
\author{C. Verdozzi$^*$, Y. Luo, and Nicholas Kioussis}
\affiliation{%
Department of Physics, California State University Nortridge, Nortridge CA 91330-8268}%
\begin{abstract}
\noindent Exact diagonalization results for Kondo nanoclusters
alloyed with mixed valence impurities show that tuning the {\it
energy spacing}, $\Delta$, drives the system from the Kondo to the
RKKY regime.  The interplay of $\Delta$ and disorder gives rise to
a $\Delta$ versus concentration $T=0$ phase diagram very rich in
structure, where regions with prevailing Kondo or RKKY
correlations alternate with domains of ferromagnetic order. The
local Kondo temperatures, $T_K$, and RKKY interactions depend
strongly on the local environment and are overall {\it enhanced}
by disorder, in contrast to the hypothesis of ``Kondo disorder''
single-impurity models.
\end{abstract}

\pacs{75.20.Hr, 75.75.+a,75.40.Cx,75.40.Mg }
\maketitle

Magnetic impurities in non magnetic hosts have been one of the
central subjects in the physics of strongly correlated systems for
the past four decades\cite{Hewson}. Such enduring, ongoing
research effort is motivated by a constant shift and increase of
scientific interest over the years, from dilute \cite{1960} to
concentrated impurities \cite{Steglich}, from periodic
\cite{UedaRMP} to disordered samples \cite{Miranda, CastroNeto},
and from macroscopic \cite{Stewart} to nanoscale phenomena
\cite{Schlottman}.

For macroscopic strongly correlated {\it d}- or  {\it f}- systems,
the central problem is understanding the variety of magnetic
phases observed experimentally upon varying the pressure
and species concentration, often giving rise to
non-Fermi-liquid (NFL) behavior at low temperatures \cite{Miranda, CastroNeto,
Stewart, Riseborough}. Various disorder-driven models have been
proposed to explain the experimentally
observed\cite{Stewart} NFL behavior\cite{Miranda, CastroNeto,
Stewart,Riseborough}. The phenomenological ``Kondo disorder''
approaches \cite{Miranda, Bernal}, based on single-impurity
models,
assume  a distribution $P(T_K)$ of Kondo temperatures $T_K$ =
exp(-1/$\rho(E_F)J$, caused by a distribution of either
$f-$electron-$c$-electron hybridization or of impurity energy
levels. These models rely  on the presence of certain sites with
very low $T_K$ spins leading to a NFL behavior at low $T$. An open
issue in such single-site methods is whether the inclusion of RKKY
interactions would renormalize and eliminate the low-$T_K$ spins.
An alternative view is the formation of large but finite magnetic 
clusters (Griffith phases) within the disordered phase through the 
competition between the RKKY and Kondo interactions
\cite{CastroNeto, Miranda2001}. The
first interaction being responsible for the quenching of the local
$f$-moment (LM)  via the screening of the conduction electrons,
whereas the latter being responsible for magnetic ordering.

On the other hand, the relevance of small strongly correlated systems
to quantum computation
requires understanding how the infinite-size properties become
modified at the nanoscale, due to the  finite energy spacing
$\Delta$ in the conduction band
\cite{Schlottman,KondoBox,Hu,Balseiro,Affleck}. Recent
advances in nanotechnology have made possible experiments in
extremely small samples, stimulating  a resurgence of interest in
the single-impurity Kondo physics at the nanoscale.  For such
small systems, controlling $T_K$ upon varying 
$\Delta$ is acquiring increasing importance since
it allows to tune the cluster magnetic behavior and
to encode quantum information. While the effect of
size or energy spacing on the single-impurity Anderson or Kondo
model has received considerable attention recently
\cite{Schlottman,KondoBox,Hu,Balseiro,Affleck},their role on {\it
dense} strongly correlated clusters with or without disorder
remains an unexplored area thus far.

In this work we present exact diagonalization
calculations for $d$- or $f$-electron nanoclusters to study the
effect of disorder and energy spacing on the interplay between the
Kondo and RKKY interactions.  While it is well known that the
cluster properties depend on cluster geometry and
size\cite{Pastor}, the motivation of the present calculations is
that they treat exactly the Kondo and RKKY interactions and they
provide a distribution of local $T_K$'s as renormalized by the
presence of intersite {\it f-f} interactions. Our results show that i)
tuning $\Delta$ can drive the nanocluster from the Kondo to the
RKKY regime, i.e. a Doniach phase diagram \cite{Doniach} in
small clusters; ii) the $\Delta$ versus alloy concentration $T=0$
phase diagram exhibits regions with prevailing Kondo or RKKY
correlations alternating with domains of ferromagnetic (FM) order; and
iii) the local $T_K$'s and the nearest-neighbor RKKY interactions
depend strongly on the local environment and are overall {\it
enhanced} by disorder. This disorder-induced enhancement of $T_K$
in the clusters is in contrast to the hypothesis of ``Kondo
disorder'' models for extended systems.


We consider a random binary alloy cluster, $A_{N-x}B_x$, of N=6
sites and different number of B atoms, $x = $ 0-N, arranged in a
ring described by the half-filled ($N_{el}=12$) two-band lattice
Anderson Hamiltonian; in standard notation,
\begin{eqnarray}
H &=& t\sum_{ij\sigma}c^{\dagger}_{i\sigma}c_{j\sigma}
+\sum_{i\sigma}\epsilon^i_{f}f^{\dagger}_{i\sigma}f_{i\sigma}
+\sum_{_i}U_{i}n_{i+}n_{i-}\nonumber\\
 & &+\sum_{i\sigma}V(f^{\dagger}_{i\sigma}c_{i\sigma}
+c^{\dagger}_{i\sigma}f_{i\sigma}).
\end{eqnarray}
\noindent {\it A. Effect of disorder}

\noindent We introduce binary disorder in the $f$-orbital energy $\epsilon^i_f$ ($\epsilon_f^A$ or
$\epsilon_f^B$) and in the intra-atomic Coulomb energy $U$ ($=U_A$ or $U_B$), to model two
different types of atoms: a Kondo-type A atom with $\epsilon_f^A =
- U^A/2$= -3 (symmetric case) and a mixed-valent (MV) type B atom
with $\epsilon_B$ = -2 and $U_B$ = 1. Both types of atoms have the same on-site
hybridization $V_A= V_B = 0.25$. For $t=1$, this choice of
parameters leads to a degeneracy between the doubly-degenerate
$c$-energy levels, $\epsilon_k = -t$ and the
energy level $\epsilon_f^B +U$ of the MV atom.
Upon filling the single particle energy levels for any $x$, $N-x$
($x$) electrons fill the $\epsilon_f^A$ ($\epsilon_f^B$) levels ,
and two electrons the -2$t$ conduction energy level, with the
remaining $N-2$ electrons accommodated in the $x$+4 degenerate
states at $-t$. This in turn results in strong charge fluctuations.

The configurations for $x\leq3$ are shown in Fig. 1, left panel, along with the
value of the spin, S$_g$, of the ground-state. The A (B) atoms are
denoted by closed (open) circles, respectively. Except for the
homogenous cases ($x$=0 and $x=6$), with a S$_g$ = 0
ground state, for all $x$ there are
configurations with S$_g \neq 0$. The average occupation and
average LM for the periodic Kondo and MV lattices are $<n_f^A> = 1$,
$<(\mu_f^A)^2>$ = 0.99, and $<n_f^B>$ = 1.6, $<(\mu_f^B)^2>$ =
0.43, respectively. We carry out a detailed analysis  for $x$=1
(S$_g$ =2) to demonstrate the FM transition induced by
a single MV atom in an otherwise Kondo cluster. Studies of
extended systems have reported similar occurrence of
ferromagnetism in the MV phase\cite{Nolting,Callaway}. As
expected, the singlet ground state of the $x=0$ Kondo cluster is
characterized by n.n. anti-ferromagnetic (AF) {\it f-f} spin
correlations ($<S_f^A(i)S_f^A(i+1)>$ = - 0.58). The introduction of
a MV atom renders them ferromagnetic. Since $U_B$ is small,
the B impurity tends to remove charge from the the conduction
band, in particular from the $k$-state with $\epsilon_k = -t$,
which has large amplitude at the B site and at the opposite A site
across the ring.  Such a depletion is different for the two spin
states, thus yielding a maximum value for the f-moment of the MV
atom. The $f$-$f$ spin correlation function between the Kondo and
MV atoms are AF ($<S_f^A(i)S_f^B(i+1)>$ = - 0.23), while they are
FM among the Kondo atoms ($<S_f^A(i)S_f^A(i+1)>$ =
+0.94). 
A similar result was  recently found in {\it ab
initio} calculations\cite {MnPRL}, where introducing a
nitrogen impurity in small (1-5 atoms) Mn clusters induces
ferromagnetism via AF coupling between the N to the Mn atoms,
whilst Mn-Mn couple ferromagnetically. We find that there is a
crossover in $S_g$ from 0 $\rightarrow 1 \rightarrow  2
\rightarrow 0$ (Fig. 1, right panel) indicating a reentrant nonmagnetic transition
around $\epsilon_B=2$. This almost saturated FM $S_g=2$
domain is robust against small changes in $U_B$, $V$,
$\epsilon_A$, $U_A$, cluster size ($N=7$), and band filling
($N_{el}$ = 10) provided that the Kondo atom has a large LM.
\begin{figure}
\vspace{-0.5cm}
\includegraphics[width=3.6in]{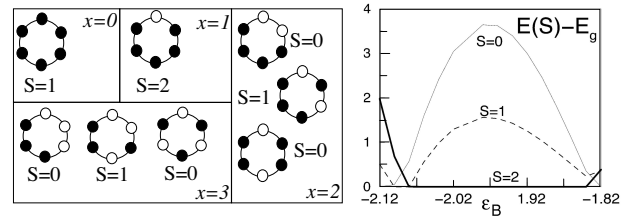}
\vspace{-4.2cm}
\caption{\label{fig:epsart}
Left panel: Alloy configurations for various concentrations $x\leq 3$ (the $x>3$ cases
are obtained by exchanging closed and open circles). For each $x\leq3$ configuration, the  
value of the ground-state spin $S_g$ is reported. Right panel: Energy difference (in
units of $10^{-4}t$ ) between the lowest $S\leq2$ eigenstates and the ground state as 
function of $\epsilon_B$.}
\end{figure}
\begin{figure}
\includegraphics[width=3.2in]{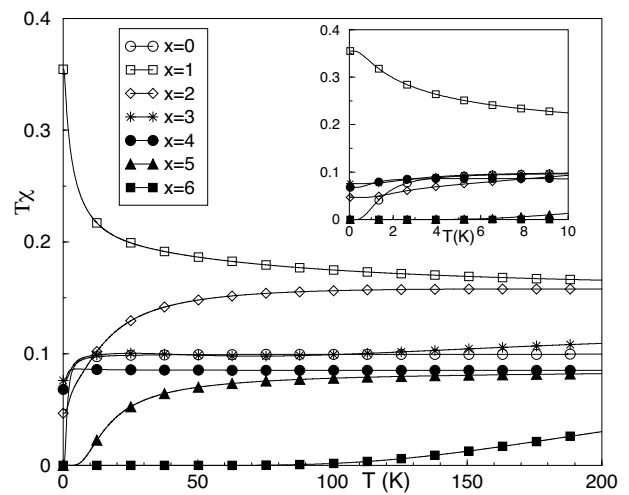}
\caption{\label{fig:epsart2}
Temperature dependence of the average f-susceptibility for
different alloy concentrations. The inset shows
the low-temperature behavior.} \protect\label{show.fig}
\end{figure}

We have also calculated the effect of disorder on the
temperature-dependent average susceptibility, $\chi^f_x(T)$,
\begin{eqnarray}
\frac{ k_BT\chi^f_x(T )}{(g\mu_B)^2}=
\sum_{C_x,\alpha_{C_x}}W^{C_x}_{\alpha_{C_x}}
e^{-\frac{E_{\alpha_{C_x}}}{k_BT}}<S_f(i)S^{Tot}>_{\alpha_{C_x}}.
\end{eqnarray}
\noindent Here, $S^{Tot}$ is the z-projection of the total spin
(both $f$- and $c$-contributions), $\alpha_{C_x}$ are the configuration-dependent exact
many-body states, and $\sum_{C\alpha} W^C_{\alpha}$ denotes exact configuration
averages. In Fig. 2 we present $T\chi^f_x(T)$
as a function of temperature for different $x$. As $T\rightarrow
0$ (inset Fig. 2) $T\chi^f_x(T)$ approaches a finite value for
$x=1-4$ while it vanishes exponentially for $x$=0, 5 and 6. This
is due to the fact that the former concentrations involve some
configurations which are magnetic, while the latter have singlet
ground states (Fig. 1). The stronger (weaker) low-temperature 
dependence for $x=1$
($x=2-4$) is due to the smaller (larger) spin gap between the
ground state and the lowest excited states. The magnetic
susceptibility displays also a magnetic crossover upon varying
$x$, and reveals a Curie-like divergence at low T
for $x=1-4$. The temperature-dependent results for the specific
heat, not reported here, show corroborative evidence of this
disorder-induced magnetic crossover.\newline

\noindent {\it B. RKKY versus Kondo: Effect of energy spacing}

\noindent Next we address a number of important open issues, namely
 (1) the effect of  $\Delta$ on the interplay between  
 RKKY and Kondo interactions in disordered 
 clusters, (2) the characterization of  the single-impurity "Kondo
correlation energy" $T_K$ in a {\it dense-impurity} cluster and
(3) the effect of disorder and $\Delta$ on the
distribution of the local $T_K$'s.  In the following, $\epsilon_B=-2$. 

In contrast with previous studies, which
introduced a phenomenological distribution $P(T_K)$ of
single-impurity Kondo temperatures, the advantage of the present
calculations is that one calculates exactly the Kondo
correlation energy: we employ  the so-called ``hybridization''
approach\cite{fulde}, with $T_K$ defined as
\begin{equation}
k_B T_K(i) = E_g(V_i=0)-E_g,
\end{equation}
where $E_g(V_i=0)$ is the ground-state energy of the
dense-impurity
cluster when V is set to zero at the i{\it th} site. Eq.(3)
reduces to $ k_BT_K =
E_{band}-E_F+\epsilon_f- E_g $\cite{Yoshida,Hu} in the
single impurity case. Here, $E_F$ is the highest occupied
energy level in the conduction band and $E_{band}$ is the
conduction band energy. This
definition of the local $T_K$ takes into account the interaction
of the $f$-moment at site $i$ with the other $f$-moments in the
system \cite{Wilkins}.
%


In Table I  we list for the periodic, $x$=0, case the local Kondo
{\it f-c} spin correlation function $<S_f^A(i)S_c^A(i)>$, the n.n. {\it f-f}
spin correlation function $<S_f^A(i)S_f^A(i+1)>$, and the local
Kondo temperature for two different values of $t$ (The energy
spacing is $\Delta=4t/(N-1)\equiv 4t/5$). As $t$ or $\Delta$
decreases the {\it f-c} spin correlation function is dramatically
enhanced while the {\it f-f} correlation function becomes weaker,
indicating a transition from the RKKY to the Kondo regime. This is
also corroborated by the increase in the local T$_K(i)$. The
energy spacing affects not only the magnetic (A) atoms but the MV
atoms as well. Thus, increasing $t$  drives the B
atoms from the non-magnetic, NM ($n_f \approx 2$), to the MV  and
finally to the Kondo regime.

\begin{figure}
\includegraphics[width=3.5in]{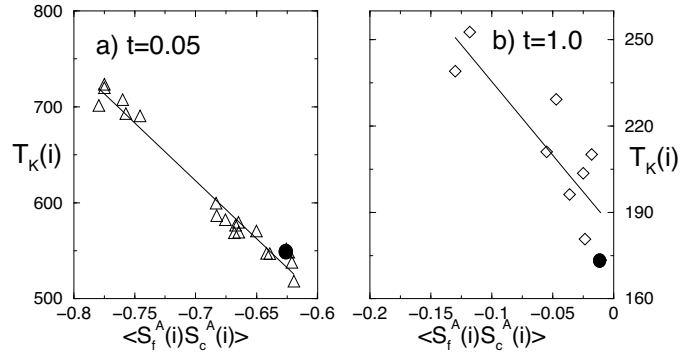}
\vspace{-0.8cm}
\caption{\label{fig:epsart} A-atoms: Local Kondo Temperatures (in K) vs the local {\it f-c} spin correlation function, for different configurations and two different values of $t$. The closed circles refer to the $x=0$ case and the lines are a guide to the eye. }
\end{figure}

\begin{table}
\caption{\label{tab:table1}  Local Kondo {\it f-c} and n.n. {\it f-f} spin
correlations functions and the local Kondo temperature (in K) for
two values of $t$ (in eV). The average energy spacing is
$\Delta=4t/(N-1)\equiv 4t/5$.}
\begin{ruledtabular}
\begin{tabular}{c|ccc}
 & $<S_f^A(i)S_c^A(i)>$ & $<S_f^A(i)S_f^A(i+1)>$ & $T_K(i) $ \\
\hline
 t=0.05  & -0.626  & -0.322  & 551.8  \\
\hline
 t=1.00  & -0.011  & -0.584  & 173.4  \\
\end{tabular}
\end{ruledtabular}
\end{table}


In Fig. 3 we present the local $T_K(i)$ as a function of the
local {\it f-c} spin correlation function  $<S_f^A(i)S_c^A(i)>$ for
all Kondo (A) atoms in the singlet ground 
state at any concentration $x$ for $t$= 0.05 and 1.0.
Note the different scales both on the horizontal and
vertical axis in the panels. In both panels, the closed circles
correspond to the $x$=0 lattice case and the line is a
guide to the eye. The results indicate a correlation between T$_K$
and the {\it f-c} spin correlation function (the larger $T_K$'s
correspond to the more negative {\it f-c} values) as one would expect,
since both provide a measure of the Kondo effect. For $t$=0.05,
most of the disordered cluster configurations are in the Kondo
regime ($S_g$ = 0), with larger $T_K$ values; consequently, panel
(a) has a larger number of singlet configurations. The
introduction of MV impurities induces a distribution of
$T_K(i)$'s, whose values are overall {\it enhanced} compared
to those for the $x$=0 case, except for several configurations for $t$=0.05,
in contrast with single-site theories for extended systems\cite{Miranda}.
It is interesting that  $P(T_K)$ for $t$=0.05 exhibits a bimodal
behavior centered about 710 and 570K, respectively: The higher
$T_K$'s originate from isolated Kondo atoms which have MV
atoms as n.n. so that the local screening of the magnetic moment
of the A atom is enhanced.
%

The effect of alloying  and $\Delta$ on the
RKKY versus Kondo competition for a given $x$
is seen in Fig. 4 (left panel), where  the configuration averaged local 
$<S_f^A(i)S_c^A(i)>_x$
and $<S_f^A(i)S_f^A(i+1)>_x$ correlation functions are plotted
as a function of $t$. The solid curves denote the uniform $x$=0
case, where we drive the cluster from the RKKY to the Kondo regime
as we decrease $t$. We find that the stronger the average Kondo
correlations are the weaker the average RKKY interactions and vice
versa. In the weak Kondo regime the configurations exhibit a wider
distribution of RKKY interactions indicating that they are sensitive to the local
environment. In contrast, in the strong Kondo regime, the Kondo (A) atoms
become locked into local Kondo singlets and the n.n. RKKY interactions are
insensitive to the local environment. Interestingly,
both energy spacing and disorder lead to an overall
enhancement of the RKKY interactions compared to the homogenous state.

In the right panel of Fig. 4 we present the  $t$ versus $x$ phase
diagram for the nanocluster at $T=0$ . We compare the
$<S_f^A(i)S_c^A(i)>_x$  and  $<S_f^A(i)S_f^A(i+1)>_x$
to assign a state of specific concentration to the Kondo or RKKY
regimes (black and gray circles, respectively), in analogy with the $x=0$ case (Table I) and with
mean field treatments\cite{lacroix}.
The horizontal gray stripes denote qualitatively ranges of $t$ where the B atoms
exhibit NM, MV and LM behavior. An interesting feature of the phase diagram is the appearance of a large FM region ($S_g \neq 0 $) enclosed by the dashed line.
The RKKY region at large $t$ and large $x$ originates from the B
atoms which become magnetic.
For the non FM configurations and for
$x<5$ the  Kondo (RKKY) correlations of the A atoms dominate at small (large) $t$, in analogy with the $x=0$ case.
On the other hand, for $x=5$ the local Kondo
correlations of the single A atom at low $t$ dominate
over the {\it f-f} correlations between the A-B and B-B pairs.
For the uniform ($x$=6) MV case we include only results in the
large $t$ regime, where the MV atoms acquire LM's which
couple antiferromagnetically.
Overall, the RKKY interactions prevail for
any concentration when $t$ is comparable or larger than the hybridization $V$.
 \begin{figure}
\vspace{-.5cm}
\hspace{-0.2cm}
\includegraphics[width=3.7in]{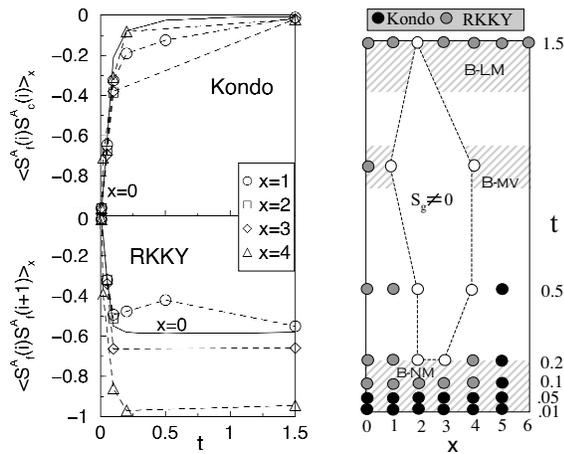}
\vspace{-1.2cm}
\caption{\label{fig:epsart}
Left panel: Configuration-averaged local {\it f-c} (top) and n.n.
{\it f-f} spin correlations (bottom) for the A atoms as function of $t$.
The solid line refers to the homogenous $x=0$ case.
Right panel: Zero-temperature $t$ vs $x$ phase diagram for the
nanocluster. Black (gray) circles denote the Kondo
(RKKY) regime. The white circles and the dashed contour
delimit the FM region. The horizontal stripes
denote the  non-magnetic (NM), 
mixed valence (MV)  and local moment (LM) behavior of the B-atoms.
}
\end{figure}
%


In conclusion, we have presented exact diagonalization
results for strongly correlated nanoclusters to study the effect of disorder and energy spacing on the interplay between the Kondo and RKKY interactions. Tuning $\Delta$  can drive the nanocluster from the Kondo to the RKKY regime,
i.e. a tunable Doniach phase diagram in small clusters. The interplay
of $\Delta$ and disorder produces a rich structure zero-temperature alloy phase diagram,
where regions with prevailing Kondo or RKKY correlations alternate with
domains of FM order. The distribution of local $T_K$ and RKKY interactions depends strongly on the local environment and are overall {\it enhanced} by disorder,
in contrast to the hypothesis of single-impurity based ``Kondo disorder'' models for
extended systems. The $\Delta$ versus disorder interplay may be relevant to experimental realizations of small cluster with tunable magnetic properties.

We acknowledge useful discussions with P. Fulde, P. Schlottmann,
P. Riseborough, A.H. Castro Neto, P.Cornaglia and C. Balseiro.  The research was supported through the NSF under Grant Nos. DMR-0097187 and  DMR-0011656 and the Keck and Parsons Foundations grants.

{}

\begin{thebibliography}{}
\bibitem{Wilkins} We also employed a second approach,
$k_BT'_K = \mu_iB_c$, where $B_c$ is
the critical local external magnetic field  necessary to break up
the singlet bound state \cite{Balseiro}. Comparative results of the methods will
be presented elsewhere.
\end{thebibliography}
\end{document}